\documentclass[conference]{IEEEtran}
\IEEEoverridecommandlockouts
\usepackage{cite}
\usepackage{amsmath,amssymb,amsfonts}
\usepackage{algorithmic}
\usepackage{graphicx}
\usepackage{textcomp}
\usepackage{blindtext, graphicx}
\usepackage{filecontents}
\usepackage{moredefs}
\usepackage{etex}
\usepackage{amsmath,amssymb,amsthm,mathtools, placeins,multirow,booktabs,rotating,graphicx,color}
\usepackage[linesnumbered,algoruled,vlined]{algorithm2e}
\usepackage{xfrac}
\usepackage{enumitem}
\usepackage{subfig}
\usepackage{array}

\usepackage{listings,multicol}
\usepackage[justification=centering]{caption}
\usepackage{multirow}
\usepackage{authblk}

\makeatletter
\newcommand{\removelatexerror}{\let\@latex@error\@gobble}
\makeatother  

\def\BibTeX{{\rm B\kern-.05em{\sc i\kern-.025em b}\kern-.08em
    T\kern-.1667em\lower.7ex\hbox{E}\kern-.125emX}}
\begin{document}

\title{Practical Sliding Window Recoder: Design, Analysis, and Usecases\\
\thanks{This is a slightly modified and extended version of a paper that will be presented at the IEEE LANMAN 2023. This work is supported by the SNOB-5G project under the MIT Portugal Program and by Northrop Grumman Corporation (NGC).}
}

\author[1]{Vipindev Adat Vasudevan}
\author[2]{Tarun Soni}
\author[1]{Muriel M{\'e}dard}

\affil[1]{Massachusetts Institute of Technology, Cambridge, MA, USA}
\affil[ ]{\textit {\{vipindev,medard\}@mit.edu}}
\affil[2]{Northrop Grumman Corporation, USA}
\affil[ ]{\textit {Tarun.Soni@ngc.com}}

\maketitle
\IEEEpubidadjcol

\begin{abstract}
Network coding has been widely used as a technology to ensure efficient and reliable communication. The ability to recode packets at the intermediate nodes is a major benefit of network coding implementations. This allows the intermediate nodes to choose a different code rate and fine-tune the outgoing transmission to the channel conditions, decoupling the requirement for the source node to compensate for cumulative losses over a multi-hop network. Block network coding solutions already have practical recoders but an on-the-fly recoder for sliding window network coding has not been studied in detail. In this paper, we present the implementation details of a practical recoder for sliding window network coding for the first time along with a comprehensive performance analysis of a multi-hop network using the recoder. The sliding window recoder ensures that the network performs closest to its capacity and that each node can use its outgoing links efficiently.
\end{abstract}

%
\section{Introduction} \label{1}
One of the key advantages of network coding \cite{ahlswede2000network} is its ability to mitigate the effects of packet loss, which can frequently occur in wireless and congested networks. Further, it allows the intermediate nodes to \textquotesingle recode\textquotesingle, mix the packets in the incoming links and send them over on the outgoing links. In Random Linear Network Coding (RLNC) \cite{ho2006random}, which is one of the most popular and widely used network coding techniques, each coded packet is a combination of multiple packets mixed together with random coefficients and the receiving node can decode the packets once it has enough innovative coded packets. Having a recoder reduces the number of transmissions required since the source can stop transmitting when the recoder has enough packets to regenerate more coded packets, thus achieving min-cut capacity. The losses in a multihop network are not cumulative anymore and each node has to compensate only for the losses in its outgoing links. Thus in a multi-hop network, network coding further improves the performance by reducing the need for retransmissions and improving end-to-end goodput and in-order delivery delay. 

Practical block RLNC implementations already have efficient recoder modules proposed and implemented efficiently. However, more advanced network coding techniques such as a sliding window network coding (SWNC) \cite{karafillis2013algorithm,wunderlich2017caterpillar} scheme, to the best of our knowledge, have no efficient recoder modules implemented. This is partially due to the difficulty in managing the window size over multiple hops and partially due to the fact that an on-the-fly block recoder module can be reconfigured to work as a simple relay module for the incoming traffic. However, such a scheme can have no significant improvement in performance rather than ensuring the availability of innovative packets at the receiver. Further, such systems will require the source node to encode the packets at a rate that can compensate for the cumulative losses in the channel. Instead, an on-the-fly sliding window recoder can allow the intermediate node to recode over different window sizes and thus adapt to the different loss rates in the links. In this work, we propose the design of a novel sliding window recoder that can work efficiently over the links with varying loss rates and allow each intermediate node to code at a different rate required for their outgoing links. Comprehensive performance analysis of the recoder compared to an end-to-end network coding scenario as well as selective repeat ARQ (SR-ARQ) in a multi-hop network is discussed. Further, we present a few practical use cases and considerations in the design. Section \ref{2} describes this network model with practical aspects. Section \ref{3} discusses the proposed design followed by the implementation and results in section \ref{4} and \ref{5} respectively. Section \ref{6} discusses some of the most interesting use cases of the proposed approach. Section \ref{7} concludes the paper and discusses potential extensions of the work.

\section{Network Model}\label{2}

\begin{figure}
\centering 
\includegraphics[width= 0.95\columnwidth]{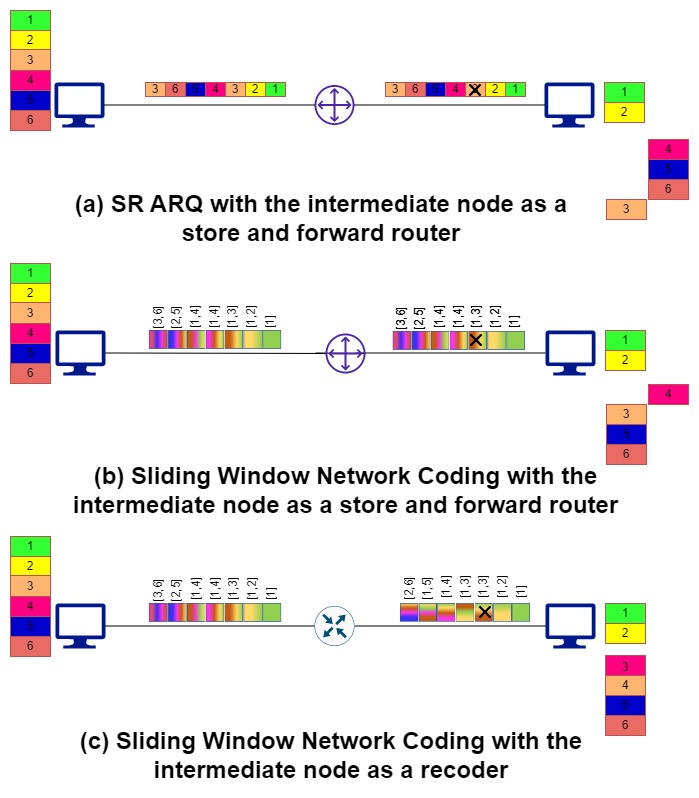}
\caption{The three different scenarios considered in this study.}
\label{fig1}
\centering
\end{figure}

Figure~\ref{fig1} shows the different network scenarios for transmitting information from a source to a receiver that we consider. The first scenario (a) is a traditional network with a source, a relay node, and a receiver. The second scenario (b) depicts an SWNC-enabled network, but only the end nodes are capable of encoding/decoding. The third case, Fig.~\ref{fig1} (c) represents the case of SWNC with a source, a recoder node, and a receiver. In this scenario, the recoder node uses linear network coding to combine packets from the source and create a new packet that is sent to the receiver. The recoder node has its own packet \textit{buffer}, which is used to temporarily store packets received from the source node. It has the ability to recode with a new code rate, to compensate for a different loss rate in its forward channel. In this paper, we will focus on the single-path scenario with only one recoder, but the recoding mechanism can be directly extended to multi-path and/or multi-hop scenarios as well.

The code rate is generally defined as the ratio of the number of information packets to the total number of packets sent, so if the sender creates $n-k$ repair packets for every $k$ innovative packets, the code rate is $(k/n)$. To address potential channel errors, it is ideal to set the code rate slightly lower than $(1-\epsilon)$, where $\epsilon$ is the expected error for the channel. The difference between $(1-\epsilon)$ and $k/n$ is referred to as the delay-goodput trade-off criterion ($\gamma$) in this case. A higher value of $\gamma$ means more repair packets will be sent, which may decrease goodput but guarantee minimum in-order delivery delay. It also ensures reliable communication, as a higher $\gamma$ value represents a higher number of repair packets to compensate for the erasures. In the sliding window, the window size is not fixed, but the introduction of repair packets is controlled by fixing the code rate. This means that the packets being coded together to create the repair packets are limited by the window size. The code rate and window size are decided considering the error probability and the round trip time in the immediate link. Considering the error probability in the first hop and the second hop as $p_{ch1}$ and $p_{ch2}$ respectively, the code rates at each node are chosen such that the redundant packets will compensate for the errors in it's forwarding channel only. This means that the code rate at the intermediate node can be different from that at the source and possibly lower than the source when $p_{ch2} > p_{ch1}$. This reduces the total number of packet transmissions required in the network as each node will only be sending enough packets such that the next hop is able to get all innovative packets. On the other hand, if we do not consider the intermediate node capable of recoding, the code rate at the source requires to compensate for the losses in the cascaded channel, $p_{combined}$. If an End-to-End coding option is chosen, without recoding at the intermediate node, the code rate can be at most $(1-p_{combined})$. However, it is evident from the min-cut max-flow principles that the true capacity of the network is higher, at $min(1-p_{ch1},1-p_{ch2})$. This is achievable with a recoder, instead of a repeater, at the intermediate node. This shows immediate improvement in the performance using the sliding window recoder at the intermediate node.

\section{System design} \label{3}

The focus of this work is on the design of a sliding window recoder, capable of adapting to different channel rates at its incoming and outgoing channels and compensating for the losses in the outgoing channels without incurring any additional retransmissions to the sources of its incoming links. A sliding window random linear network coding (SW RLNC) is employed to achieve low latency and reliable communication in the network. A practical implementation of a sliding window recoder has never been studied in the literature. The SW RLNC operates with a fixed code rate, which means that after a certain number of new packets are sent, a set number of Forward Erasure Correction (FEC) packets will be transmitted as well. The recoder allows each node to set its code rate to account for errors on its outgoing link. In exceptional cases of extreme error bursts where apriori repair packets are insufficient, feedback-based FEC can be utilized. This may reduce the code rate slightly but ensures all packets are transmitted reliably.

The sliding window at the sender and receiver end is very similar to the base description in the literature \cite{cloud2015network}. However, at the intermediate node, it is important to consider that the packets sent out of it have innovative packets capable of overcoming the extra losses at its outgoing channel. The recoder should be \textquotesingle carefully\textquotesingle~mixing the packets and the original window sizes are maintained on the forward packets. An example scenario is presented in Fig. \ref{fig_example} and explained step by step in the appendix for more details.

\begin{figure}
\centering 
\includegraphics[width= 0.95\columnwidth]{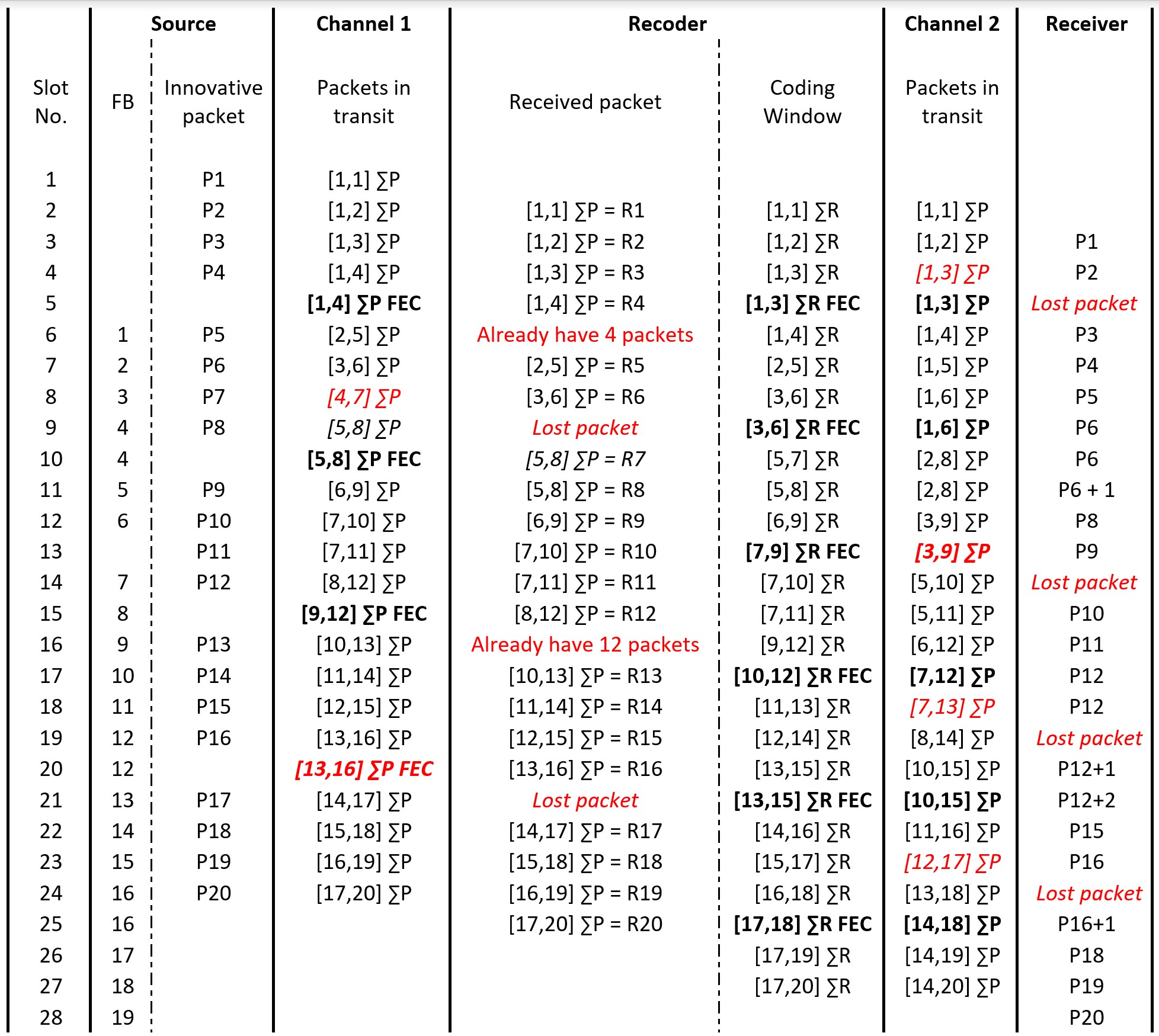}
\caption{An example case of a 2-hop network with SWNC. The red-colored transmissions are lost in the channel.}
\label{fig_example}
\centering
\end{figure}

\subsection{Recoding Process}
The packets received from incoming links are stored in a buffer along with their respective window information. At the slot for sending a packet in the forward channel, the recoder selects the window size from available packets and combines them into a single packet. The recoder starts by selecting the first packet in the window, adding it to the recoding set, and using its opening window slot as the opening window slot of the outgoing packet. The recoder continues to add packets until the desired window size is reached. The closing window of the last included packet becomes the closing window value of the recoded packet. It is important to note that the window size at the recoder may not match the difference between the end values of the window associated with the recoded packet, as the window is maintained based on the windows of the incoming packets.

The next concern at the recoder after managing the window properly is to handle the coefficients. The outgoing packets should have coefficients that will allow the decoder to decode the packet. This is accomplished by adding the coefficients to the packet payload. The recoder performs the coding operation on both the packet payload and the incoming coefficients. Once the decoder has enough packets, it separates the received coefficients, calculates their inverse matrix, and decodes the message.  This results in an increase in the payload size that consists of the actual payload length and the number of added coefficients, so a small reduction in the maximum payload size that can be used at the source node. However, it is important to note that the number of coefficients does not increase as the packet travel through the network and is bounded by the maximum window size at the encoder. The recoding algorithm is presented in alg.~\ref{algorithm1}.

\begin{algorithm}[b]
\caption{Recoder's Process}
\label{algorithm1}
\DontPrintSemicolon
\LinesNotNumbered
\KwIn{Incoming packet and coding headers}
\KwOut{Outgoing packet and coding headers}
$\mathbf{Step\ 1}$:
Identify min and max window values and coded payload \\
$\mathbf{Step\ 2}$: \eIf{the incoming packet is innovative}{Add the incoming packet to buffer \;
}{
Discard the packet \;
}
$\mathbf{Step\ 3}$: \If{slot to send a new packet}{Add the next packet in buffer to the window} 
$\mathbf{Step\ 4}$: Generate coefficients for the current window and create a recoded packet \;
$\mathbf{Step\ 5}$:
Add the min window of first packet and the max window of last packet in the recoding window as the min and max of the outgoing packet's window\\
$\mathbf{Step\ 6}$:
Send the outgoing packet with window limits
\end{algorithm}

\subsection{Feedback Mechanism}
Network coding-based approaches do not rely on the arrival order or sequence number of packets. Instead, any coded packet can replace a lost packet within its coding window. The source node uses feedback from the destination to move the window forward, taking into account the number of innovative packets received at the decoder. This feedback is designed such that it includes both fully decoded packets and partially decoded packets in the coding window that still need additional packets for complete decoding.

The window at each node is advanced based on the feedback received. The feedback value is compared to the window closing value of the packets, and any packet that does not contain an innovative packet (i.e., a packet with a value higher than the feedback value) is removed from the coding window. If there is a large burst of errors, the window will continue to grow until it reaches its maximum capacity. Different decision-making criteria can be used if the window reaches its capacity. For example, if every packet is critical, the window can be kept at that length and more repair packets can be sent. However, in many practical scenarios, the freshness of the packet is a crucial factor, such as in live video streaming where some lost packets may not be worth retransmitting. In such cases, the oldest packets may be dropped from the window to keep it sliding \cite{cloud2015network}.

\section{Implementation}  \label{4}
The KODO library by Steinwurf ApS \cite{kodo} optimizes network coding implementations. Although it has been one of the major building blocks of network coding implementations and is available in various programming languages, a straightforward sliding window recoder is not yet available within KODO. This work modifies the sliding window APIs in the library to create a recoder, which was not a feature previously considered even in SWNC standardization efforts \cite{roca2020rfc}. The first step was to define the coding header and payload structures. As they are common in packet networks, headers often include additional information about the message such as source and destination IPs and flags. In network coding implementations, the header also includes information necessary for decoding the message. This section also discusses the network coding-related header values and flags.

In a sliding window mechanism, the size and opening point of the window are crucial factors and are therefore included in the coding header. The size of the window is represented by 1 byte, while the opening point size is represented by 2 bytes. The number of coefficients is limited by the window size, which is also included in the coding header (1 byte). To distinguish between repair packets and packets with innovative data compared to the previous packet, a single bit, 0 or 1, is added. In multi-hop systems, two flags are used to distinguish repair packets: the \textquotesingle source FEC flag\textquotesingle, which is set to 1 by the source node when generating a repair packet, and the \textquotesingle  last FEC flag\textquotesingle, which is changed by each node and set to 1 when a node generates a repair packet. Both flags are set to 1 in the source node but may be different in other nodes. The coding header is 5 bytes, with the last 6 bits kept null.

The essence of network coding is the possibility of creating new packets from a combination of original packets using random coefficients. These coefficients are necessary for decoding at the receiver; so they are either attached to the packet payload or header. The size of the coefficients is determined by the number of packets included in the coding window. In end-to-end coding schemes, where coding is performed only at the source and decoding occurs at the sink, the coefficients can be shared by sharing the seed value used to generate them, along with the window size, instead of sharing the coefficients themselves.

However, when recoding is performed, the coefficients must reflect the corresponding operations. Simply sharing the seed and recreating the original coefficients is not ideal for a network that wants to use network coding optimally by recoding at intermediate nodes. To ensure that recoding operations are reflected in the coding coefficients, we propose a simple change to the packet payload. The source node performs the encoding and attaches the coding coefficients to the payload, resulting in a payload size slightly larger than the original packet size. This simplifies the recoding process for nodes, as they can simply recode the packet payload using locally generated coefficients. The sink node can then disassemble the original payload and coefficients and decode correctly using the received coefficients. It's worth mentioning that the size of the payload does not increase as it travels through the network. The payload length is determined only by the maximum coding window and recoder nodes simply perform the recoding without adding their own coefficients to the payload.

The feedback packet is also defined slightly differently from a simple TCP-style packet, with a minor modification. Instead of sending the sequence number of the last correctly received packet, our approach requires the receiver to send two values: the number of correctly decoded packets and the number of partially decoded packets. Partially decoded packets refer to packets with innovative information that have not yet been fully decoded because there are not enough packets available compared to the window size of the arrived packet. These packets can be completely decoded as soon as a repair packet arrives that completes the window for decoding. This is a critical aspect of network coding that reduces the delay in in-order delivery.

\section{Analysis and results}  \label{5}
The sliding API of the KODO-Python library was utilized to create a recoder, which was then tested using a multi-hop network to verify that the intermediate nodes could perform recoding operations. The recoder's performance was compared to an end-to-end SWNC approach. The channels are modeled as binary erasure channels. We expect the feedback channels to be error-free. The study also incorporated an SR-ARQ implementation as the baseline scenario. These three distinct scenarios were examined to evaluate the approach's efficacy. In the end-to-end network coding scenario, the cumulative channel loss was calculated, and the code rate was adjusted accordingly. We ran trials by sending 100 packets of 100 bytes each over a 2-hop network, with different error rates and RTTs fixed in the simulation. 

Three performance metrics are being assessed: completion time of information exchange, number of transmissions required, and the success ratio of packet transmission. The completion time of information exchange is calculated by measuring the number of time slots needed to receive all the packets at the receiver and acknowledge their receipt at the source. This corresponds to the active time of the network and resource utilization. The number of transmissions is the total number of forward transmissions that occur throughout the network. This metric can be related to the transmission energy spent while the information exchange. The success ratio is measured as an indicator of the user’s quality of experience. In the scope of this work, we define success ratio in terms of packets and time slots; as the ratio between the number of useful data packets received by the destination node to the number of slots. As a general convention, the solid lines correspond to the scenario of SWNC with recoder while the dotted lines correspond to the case of SWNC without recoder, and the dashed lines show traditional SR ARQ. 

\begin{figure}
\centering 
\includegraphics[width= 0.95\columnwidth]{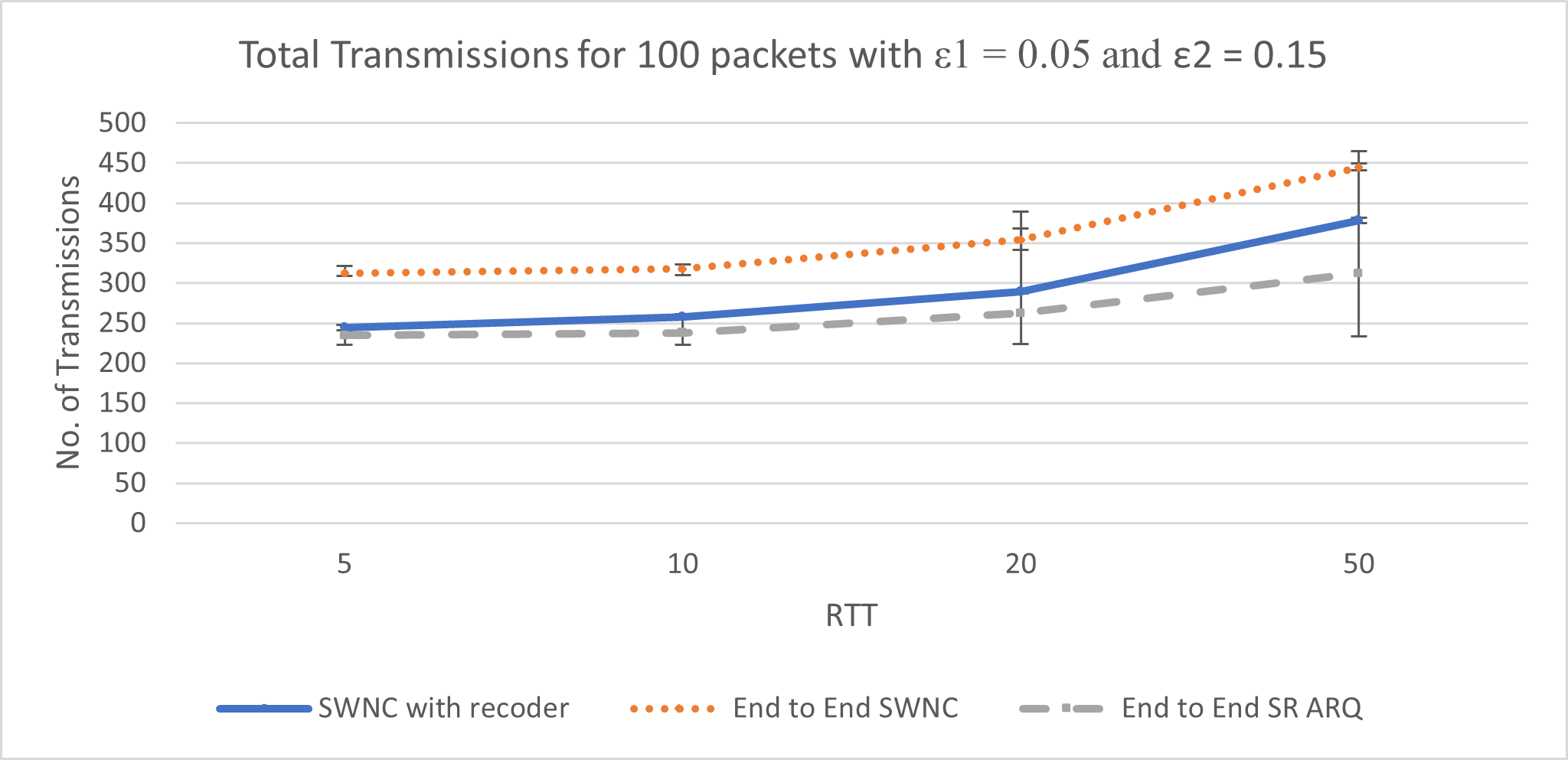}
\caption{Total transmissions till the successful reception of 100 packets when $\epsilon1 = 0.05$ and $\epsilon2 = 0.15$}
\label{fig_trans}
\centering
\end{figure}

\begin{figure}
\centering 
\includegraphics[width= 0.95\columnwidth]{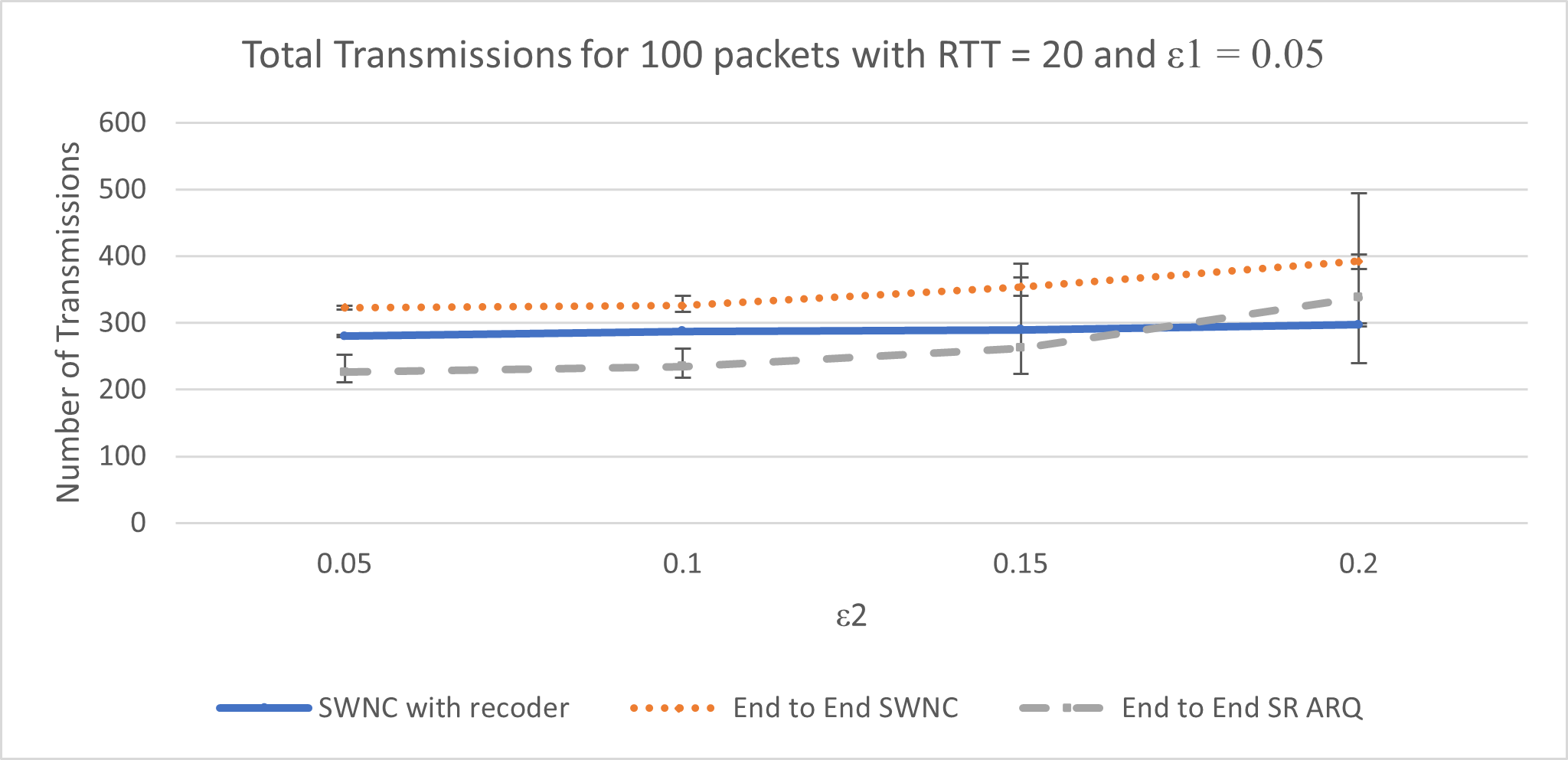}
\caption{Total transmissions till the successful reception of 100 packets when RTT is 20 slots and $\epsilon1 = 0.05$, but $\epsilon2$ varies}
\label{fig_trans1}
\centering
\end{figure}

\begin{figure}[t]
\centering 
\includegraphics[width= 0.95\columnwidth]{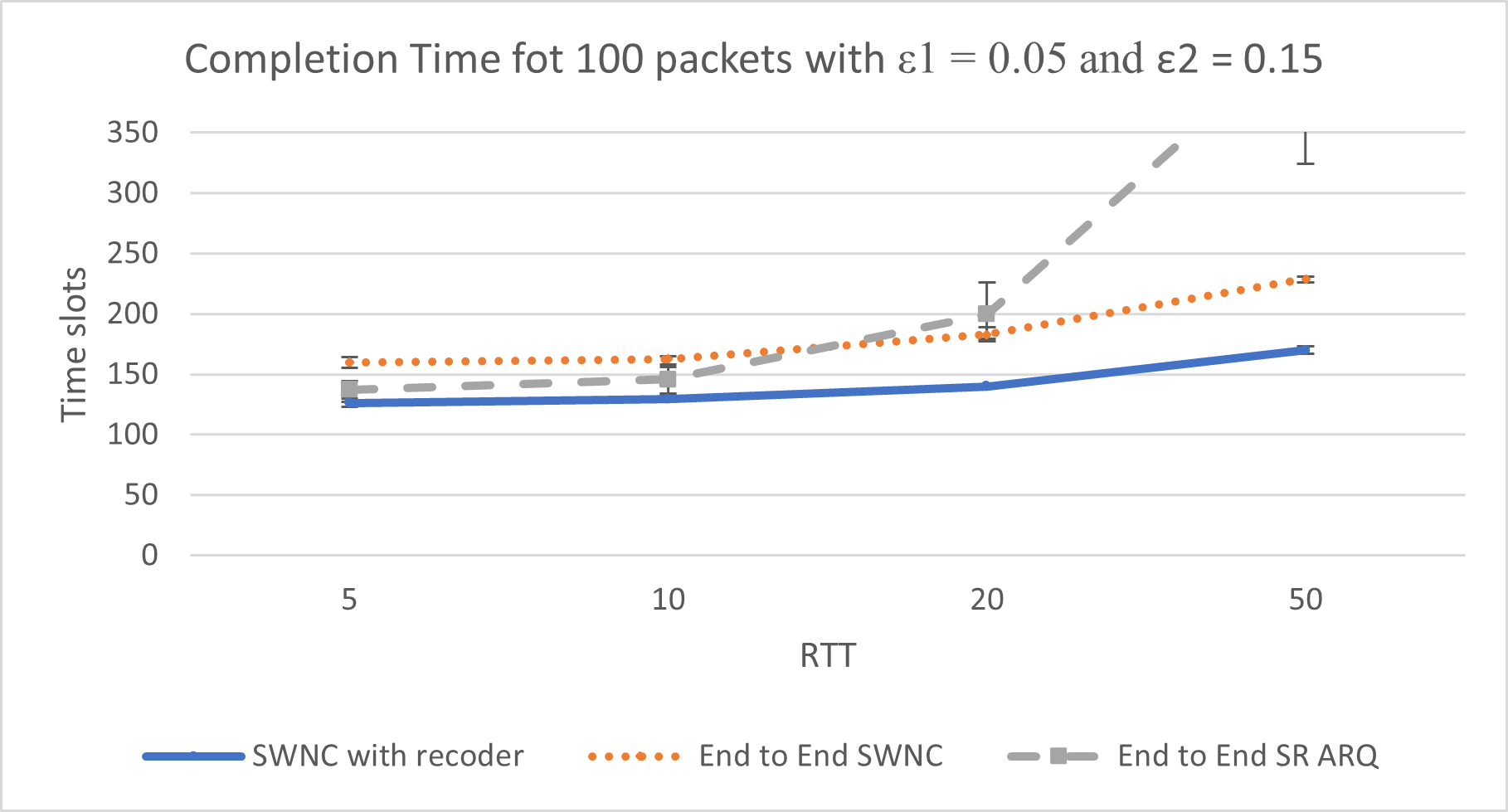}
\caption{Completion time for 100 packets when $\epsilon1 = 0.05$ and $\epsilon2 = 0.15$}
\label{fig_comp}
\centering
\end{figure}

\begin{figure}
\centering 
\includegraphics[width= 0.95\columnwidth]{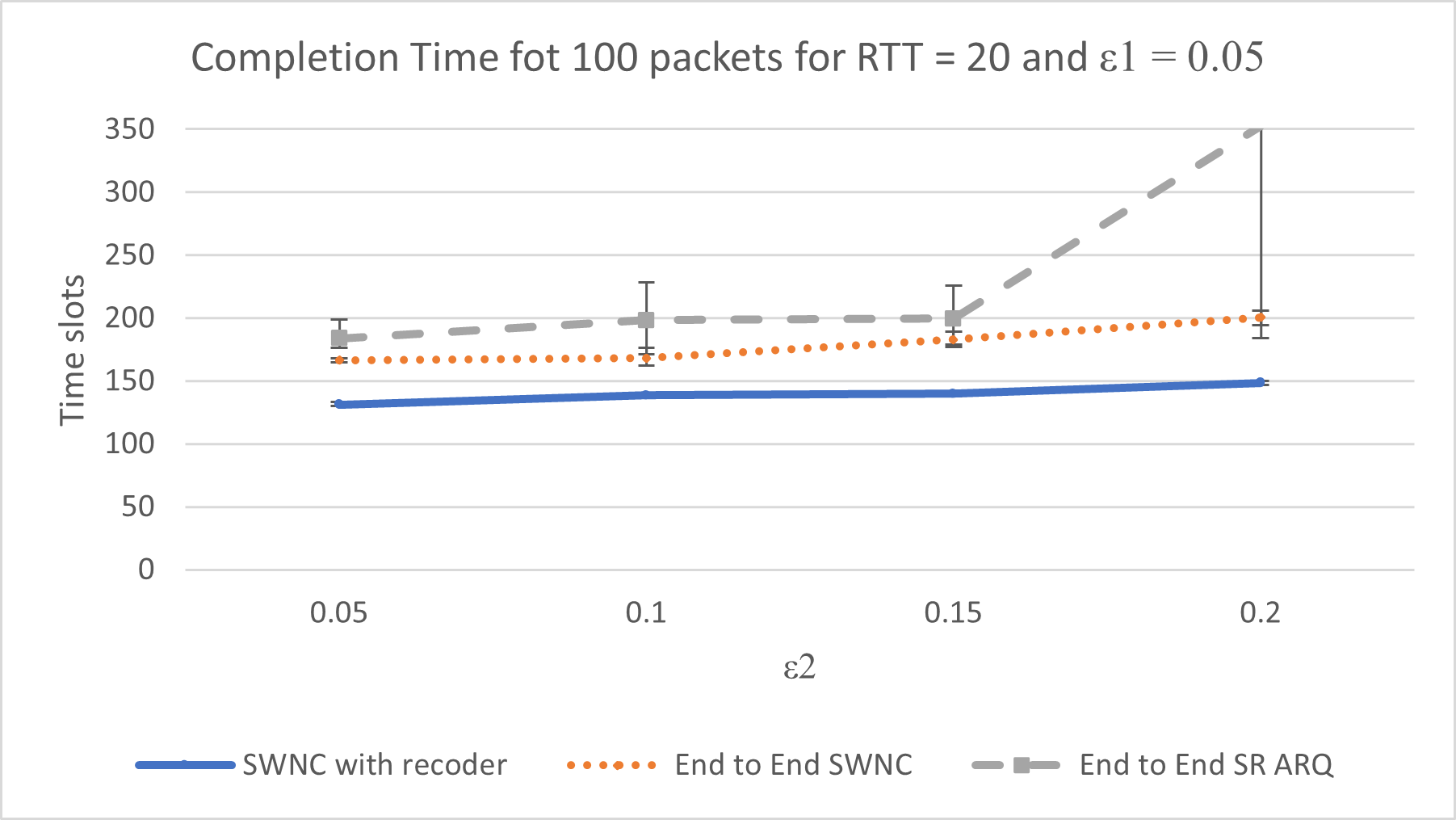}
\caption{Completion time for 100 packets when RTT is 20 slots and $\epsilon1 = 0.05$, but $\epsilon2$ varies}
\label{fig_comp1}
\centering
\end{figure}

The sliding window recoder offers a significant advantage in terms of reducing completion time in the network. As previously mentioned, the recoder enables the use of different code rates at each node, and the source can stop transmitting once the immediate intermediate node has received sufficient packets to reconstruct all the packets. This approach can also be extended to other intermediate nodes in a multi-hop network, allowing each node to halt transmissions once enough packets have been received by the next node. This considerably reduces the number of transmissions in the system compared to end-to-end SWNC, where the source and all intermediate nodes transmit duplicate packets in case of a lost packet (Fig. \ref{fig_trans} and Fig. \ref{fig_trans1}). However, SR ARQ still has the least number of transmissions unless the error rate goes much higher since the retransmissions are only happening when a NACK is received. The SWNC with recoder has a slightly higher number of transmissions because it sends more repair packets till the acknowledgments for all packets are received. However, it is to be noted that the network coding implementations are more reliable about completion as for any given RTT and loss rate, the performance can be guaranteed with very low variations. SR ARQ performance will largely depend on the loss patterns and such guarantees can not be provided. In fact, in many trials with higher RTT or loss rate, the transmission of 100 packets where not completed within 500 slots, and that is reflected in the figures \ref{fig_comp} and \ref{fig_comp1}. Furthermore, experimental trials conducted at different loss rates and round-trip times confirm that the advantage of using a recoder increases with increasing loss rate and RTT, and its performance benefits multiply as the number of hops increases.

\begin{figure}
\centering 
\includegraphics[width= 0.95\columnwidth]{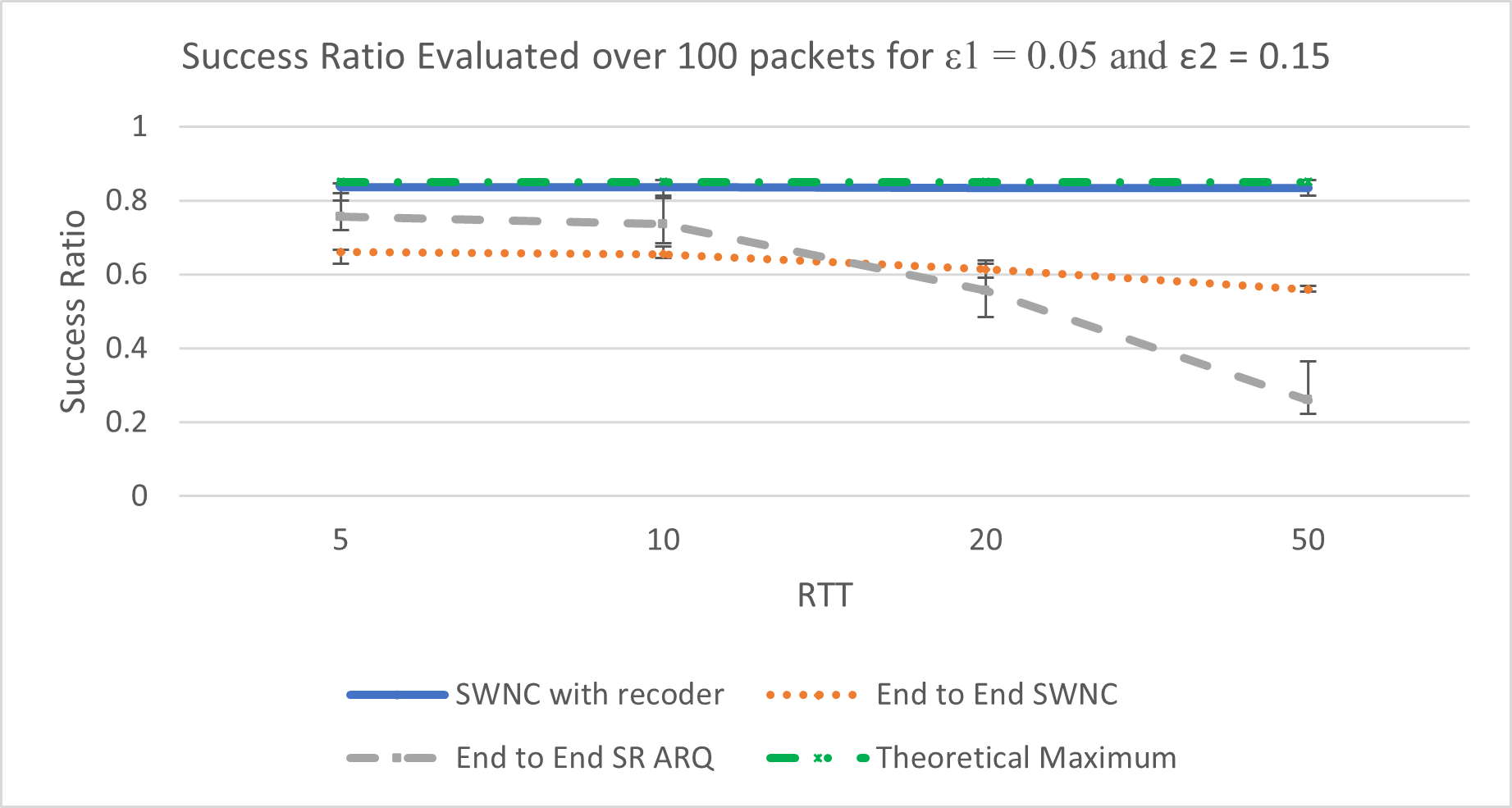}
\caption{Success ratio observed by simulation compared to the theoretical maximum for varying RTT.}
\label{fig_PDR}
\centering
\end{figure}

Practical implementations validate that the code rate achievable with the recoder is generally better than the code rate required for successful transmissions in end-to-end SWNC. When the code rate is the same, the performance remains almost unchanged, as observed by comparing instances of $\epsilon2 = 0.10$ and $\epsilon2 = 0.15$ where the same code rate is employed at the recoder, from Fig.\ref{fig_PDR}. It is worth noting that failure to employ recoding would result in the source bearing all losses between the source and destination, making it necessary to have a higher redundancy at the source. This results in reducing the success ratio since more transmissions will be required to complete a specific number of packets compared to the case with recoder. Our experimentally obtained values, as shown in Fig. \ref{fig_PDR}, exhibit a decline in the large RTT case, given that the RTT is of the same order of magnitude as the total number of packets. However, theoretical analysis of the success ratio, when the number of packets is infinite or when the number of packets is significantly higher than the RTT, reveals that it will remain unchanged for different RTTs, provided that there are enough repair packets to complete transmissions within a bounded time and negligible variance. These analyses show that using a recoder offers several advantages over an end-to-end coding solution, including a shorter bound for completion time (as seen in the completion time analysis) and improved performance on multiple fronts.

\begin{figure}
\centering 
\includegraphics[width= 0.95\columnwidth]{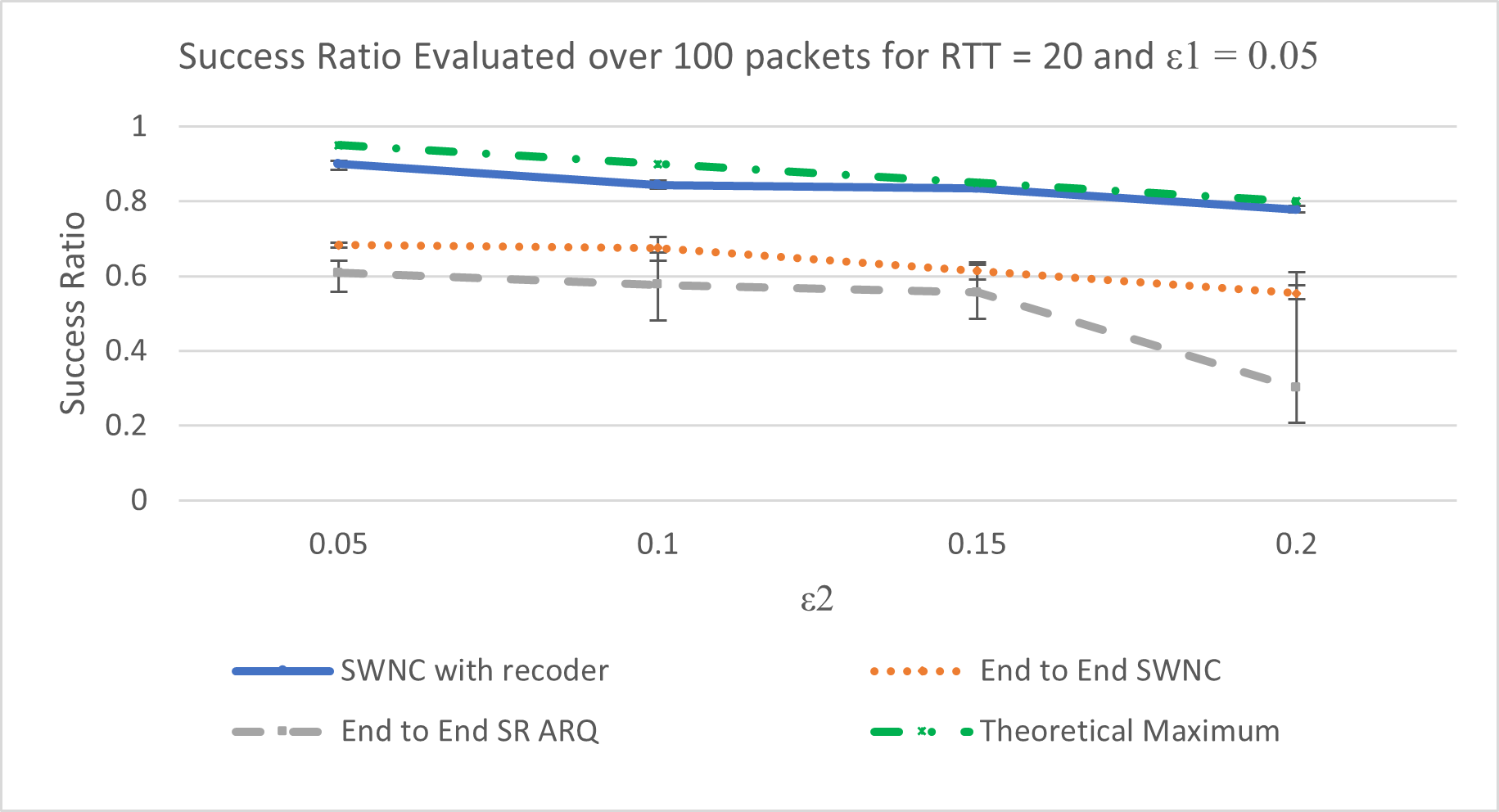}
\caption{Success ratio observed by simulation and the theoretical maximum for varying error rate and fixed RTT.}
\label{fig_PDR1}
\centering
\end{figure}


\section{Usecases} \label{6}
\subsection{Underwater Communication}
As communication networks get extended to the underwater, wireless communications started to explore different types of propagation technologies such as acoustic and visible light in order to address the differences from over-the-air RF transmission in the underwater scenario. While the acoustic-based transmission is mostly limited by the transmission rate and delay, recent underwater optical wireless communication (UOWC) trials have shown high attenuation even for a short distance communication \cite{wu2017blue}. With the terrestrial networks being transformed to the 5G standards, a low delay, high throughput, and reliable underwater communication is becoming a bottleneck in many applications. Further, the physical boundaries of the network are fading as the devices in the network spread across terrestrial and underwater locations. Different technologies such as visible light communication and laser-based communication are being tested to provide smooth single-mode communication in such networks. On the other hand, the combination of different technologies, such as RF above water level and acoustic channel underwater \cite{tonolini2018networking} are also examined to utilize the best possible situations on both sides. However, such a system demands an intermediate node capable of converting the signals from one form to another at the intersection of the two mediums. A buoy or a boat on the sea can support such hardware. However, the underwater channel still has more losses than the RF channels on air. A sliding window recoder at the buoy, instead of a conventional store and forward repeater, can bring significant improvements in the performance of the network.

\subsection{D2D Multihop Networks}
In 5G and beyond networks, Device-to-device (D2D) communication is expected to play a critical role in enabling new services and applications. One of the key benefits of D2D communication is the potential to reduce network congestion and improve spectral efficiency by enabling direct communication between user equipments without going through a base station. This can lead to reduced latency, higher throughput, and improved energy efficiency. In D2D multihop network, using a sliding window recoder can significantly improve the network's performance compared to end-to-end SWNC. The probability of packet loss increases with the number of hops in such a network and end-to-end network coding requires all nodes in the path to transmit the same packets they receive, leading to redundant transmissions and increased congestion. With a sliding window recoder, each recoder allows for different code rates to be used at each node, enabling the optimization of each link's performance. Additionally, as the number of hops increases, the network efficiency also increases with SWNC recoder. Overall, using a sliding window recoder in a D2D multi-hop network can improve the network's reliability and efficiency while reducing congestion and energy consumption.

\subsection{IoT Networks}
Another potential use case for sliding window recoder is in Internet of Things (IoT) networks. IoT networks typically consist of a large number of low-power devices with limited processing and communication capabilities. These devices often operate in harsh environments with high packet loss rates, which can result in poor network performance and reliability. Furthermore, these devices may be spread out over a large area, and may not always have a direct connection to the gateway. With a sliding window recoder, each intermediate node can perform recoding operations on the packets it receives before forwarding them to the next hop, reducing the number of transmissions required to deliver the packets to the destination. This can save energy and reduce latency, which are critical factors in IoT networks. Moreover, IoT networks often require real-time data processing and analysis, and a sliding window recoder can ensure that the packets are delivered in a timely and reliable manner. Thus, the sliding window recoder has the potential to improve the reliability, efficiency, and performance of IoT networks, making it a promising technology for the future of IoT.

\section{Conclusions} \label{7}
In this paper, we designed the sliding window recoder modifying sliding window API of KODO-Python. As far as we know, this is the first practical implementation of the sliding window recoder. The SWNC can play a crucial role in achieving ultra-reliable low-latency communication and a recoder ensures that the optimal performance over multi-hop networks is achieved. The proposed approach is tested and compared against end-to-end SWNC and SR-ARQ to showcase the benefits of the sliding window recoder. We have also provided an example scenario and various use cases of the sliding window recoder, highlighting its potential to be applied in different network scenarios. The results of this study reiterate that the recoder can significantly improve network performance and achieve the benefits of network coding at its best.



\bibliographystyle{IEEEtran}
\bibliography{references}

\section*{Appendix}
Here we are going to explain the working of recoder with the help of an example. We consider the source to have a code rate of $(4/5)$ while the recoder to have a code rate of $(3/4)$. Fig.~\ref{fig_example} shows the slots in which transmissions occur. We consider the packets to be delivered to the next hop with a delay of 1 slot while the feedback is received at the corresponding source node after 3 slots of receiving a packet. An original packet is denoted as $P{i}$ while the packets received and stored at the recoder are denoted as $R{i}$. $\Sigma P{i}$ defines the coded packet including packets up to $P{i}$. The red packets in the transition are lost before reaching their destination. Now we will explain some of the interesting incidents in the system.

1. At slot 4, the recoder sent a coded packet including the first three packets, but it gets lost in transition. However, the next packet is a repair packet from the recoder which will compensate for the lost packet.

2. At slot 6, the recoder receives a coded packet that includes original packets up to $P4$. However, the recoder already has 4 coded packets and this new packet does not provide any additional degrees of freedom for the recoder. Thus this packet is discarded at the recoder.

3. At slot 9, the packet coded over packets $P4$ to $P7$ is lost in channel 1. The next packet which includes $P8$ will be received at the next slot and will be added as the 7th packet at the recoder (marked as $R7$). Thus in slot 10, where the recoder is including $R7$ to the coding window, the outgoing packet has information up to the original packet $P8$ as you can see in the packets in transit at channel 2. This packet arrives at the receiver at slot 11 and now the receiver has 6 packets decoded completely, 1 packet partially decoded, and 1 packet missing to be completely decoded. 

4. At slot 10, the source sends a repair packet which can compensate for the loss in slot 7. This gets stored as $R8$ and included in the coding window of the recoder and helps to decode all packets up to $P8$.

5. It is to be noted that the window indicated for the outgoing packet of the recoder is the same as the minimum of the opening window of any packet and the maximum of the closing window of any packet included in its coding window. The sliding window is moved either by the feedback or when the buffer capacity of the node is reached.
\end{document}